\documentclass[preprint,footinbib,preprintnumbers,amsmath,amssymb,prb]{revtex4}

\usepackage{graphicx}
\usepackage{dcolumn}
\usepackage{bm}
\usepackage{epsfig}
\usepackage{amsmath}

\begin{document}

\preprint{Am.J.Phys./Ward}
\bibliographystyle{apsrev}

\title{How to Derive the Schr\"{o}dinger Equation}

\author{David W. Ward}
\email{david@davidward.org} \homepage{http://www.davidward.org}
\affiliation{Department of Chemistry and Chemical Biology\\Harvard
University, Cambridge, Massachusetts 02138}

\author{Sabine Volkmer}
\affiliation{Department of Physics\\Massachusetts Institute of
Technology, Cambridge, Massachusetts 02139}

\date{\today}

\begin{abstract}
We illustrate a simple derivation of the Schr\"{o}dinger equation,
which requires only knowledge of the electromagnetic wave equation
and the basics of Einstein's special theory of relativity. We do
this by extending the wave equation for classical fields to photons,
generalize to non-zero rest mass particles, and simplify using
approximations consistent with non-relativistic particles.
\end{abstract}


\maketitle

\section{Introduction}

One of the most unsatisfying aspects of any undergraduate quantum
mechanics course is perhaps the introduction of the Schr\"{o}dinger
equation. After several lectures motivating the need for quantum
mechanics by illustrating the new observations at the turn of the
twentieth century, usually the lecture begins with: ``Here is the
Schr\"{o}dinger equation.'' Sometimes, similarities to the classical
Hamiltonian are pointed out, but no effort is made to derive the
Schr\"{o}dinger equation in a physically meaningful way. This shortcoming is not remedied in the standard quantum mechanics textbooks either \cite{griffiths,merzbacher, sakurai}. Most students and professors will tell you that the Schr\"{o}dinger equation cannot be derived.  Beyond the standard approaches in modern textbooks there have been several noteworthy attempts to derive the Schr\"{o}dinger equation from different principles \cite{hall, peice, field, ogborn}, including a very compelling stochastic method \cite{nelson}, as well as useful historical expositions \cite{bernstein}.

In this paper, we illustrate a simple
derivation of the Schr\"{o}dinger equation, which requires only
knowledge of the electromagnetic wave equation and the basics of
Einstein's special theory of relativity. These prerequisites are
usually covered in courses taken prior to an undergraduate's first
course in quantum mechanics.

\section{A brief history of Quantum and Wave Mechanics}

Before we begin to derive the Schr\"{o}dinger equation, we review the physical
origins of it by putting it in its historical
context.

\begin{figure*}
\epsfig{file=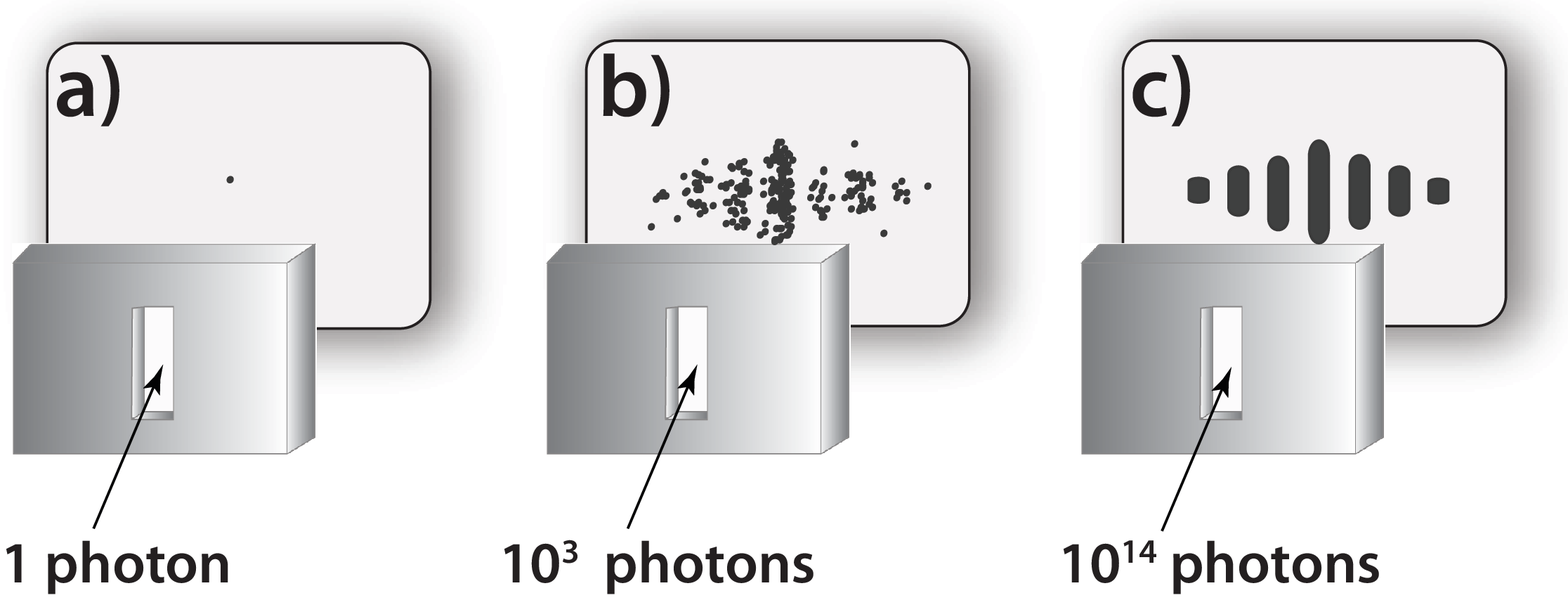,width=1.0\columnwidth}
\caption{\label{fig:slits} Interference effect in the single slit experiment for a) 1 , b) $10^{3}$, and c) $10^{14}$ photons or particles.}
\end{figure*}

The new paradigm in physics which emerged at the beginning of the
last century and is now commonly referred to as quantum mechanics
was motivated by two kinds of experimental observations: the
``lumpiness'', or quantization, of energy transfer in light-matter
interactions, and the dual wave-particle nature of both light and
matter. Max Planck could correctly calculate the spectrum of
black-body radiation in 1900 by postulating that an electromagnetic
field can exchange energy with atoms only in quanta which are the
product of the radiation frequency and the famous constant $h$,
which was later named after him \cite{planck}. Whereas for Planck himself, the
introduction of his constant was an act of desperation, solely
justified by the agreement of the calculated with the measured
spectrum, Albert Einstein took the idea serious. In his explanation
of the photoelectric effect in 1905, he considered light itself as
being composed of particles carrying a discrete energy \cite{einstein}. This bold
view was in blatant contradiction with the by then established
notion of light as an electromagnetic wave. The latter belief was
supported, for instance, by the observation of interference: If we
shine light on a single slit placed in front of a scintillating
screen, we observe a pattern of darker and brighter fringes or
rings. But what happens if Einstein's light particles, let us call
them photons, exist and we zing them one-by-one at the same slit?
Then, each photon causes the screen to scintillate only at a single
point. However, after a large number of photons pass through the
slit one at  a time, we once again obtain an interference pattern. This build-up of interference one photon at a time is illustrated in Fig.~\ref{fig:slits}.

In 1913, Niels Bohr succeeded in deriving the discrete lines of the
hydrogen spectrum with a simple atomic model in which the electron
circles the proton just as a planet orbits the sun, supplemented by
the ad-hoc assumption that the orbital angular momentum must be an
integer multiple of $\hbar=h/2\pi$, which leads to discrete
energies of the corresponding orbitals \cite{bohr}. Further,
transitions between these energy levels are accompanied by the
absorption or emission of a photon whose frequency is $E/h$,
where $E$ is the energy difference of the two levels. Apparently, the
quantization of light is strongly tied to the quantization within
matter. Inspired by his predecessors, Louis de Broglie suggested
that not only light has particle characteristics, but that classical
particles, such as electrons, have wave characteristics \cite{broglie}. He associated the wavelength $\lambda$ of these
waves with the particle momentum $p$ through the relation
$p=h/\lambda$. Interestingly, Bohr's condition for the
orbital momentum of the electron is equivalent with the demand that
the length of an orbital path of the electron has to be an
integer multiple of its de Broglie wavelength.

Erwin Schr\"{o}dinger was very intrigued by de Broglie's ideas and set
his mind on finding a wave equation for the electron. Closely
following the electromagnetic prototype of a wave equation, and
attempting to describe the electron relativistically, he
first arrived at what we today know as the Klein-Gordon-equation. To
his annoyance, however, this equation, when applied to the hydrogen
atom, did not result in energy levels consistent with Arnold
Sommerfeld's fine structure formula, a refinement of the energy
levels according to Bohr. Schr\"{o}dinger therefore retreated to the
non-relativistic case, and obtained as the non-relativistic limit to
his original equation the famous equation that now bears his name.
He published his results in a series of four papers in 1926 \cite{schroedinger1,schroedinger2,schroedinger3,schroedinger4}.
Therein, he emphasizes the analogy between electrodynamics as a wave
theory of light, which in the limit of small electromagnetic wavelength approaches
ray optics, and his wave theory of matter, which
approaches classical mechanics in the limit of small de Broglie
wavelengths. His theory was consequently called wave mechanics. In a
wave mechanical treatment of the hydrogen atom and other bound
particle systems, the quantization of energy levels followed
naturally from the boundary conditions. A year earlier, Werner
Heisenberg had developed his matrix mechanics \cite{heisenberg}, which yielded the values of all
measurable physical quantities as eigenvalues of a
matrix. Schr\"{o}dinger succeeded in showing the mathematical
equivalence of matrix and wave mechanics \cite{schroedinger5}; they
are just two different descriptions of quantum mechanics. A
relativistic equation for the electron was found by Paul Dirac in
1927 \cite{dirac}. It included the electron spin of $1/2$, a purely
quantum mechanical feature without classical analog. Schr\"{o}dinger's
original equation was taken up by Klein and Gordon, and eventually
turned out to be a relativistic equation for bosons, i.e. particles
with integer spin. In spite of its limitation to non-relativistic
particles, and initial rejection from Heisenberg and colleagues, the
Schr\"{o}dinger equation became eventually very popular. Today, it
provides the material for a large fraction of most introductory
quantum mechanics courses.

\section{The Schr\"{o}dinger equation derived}

Our approach to the Schr\"{o}dinger equation will be similar to that
taken by Schr\"{o}dinger himself. We start with the classical wave
equation, as derived from Maxwell's equations governing classical
electrodynamics (see the appendix). For simplicity, we consider only
one dimension,
\begin{equation}
\frac{\partial^2E}{\partial^2x}-\frac{1}{c^2}\frac{\partial^2E}{\partial^2t}=0.\label{maxwellwaveeqn}
\end{equation}
This equation is satisfied by plane wave solutions,
\begin{equation}
E(x,t)=E_0e^{i(kx-\omega t)},\label{planewavesoln}
\end{equation}
where $k=2\pi/\lambda$ and $\omega=2\pi\nu$ are the spatial and
temporal frequencies, respectively, which must satisfy the
dispersion relation obtained upon substitution of
Eq.~(\ref{planewavesoln}) into Eq.~(\ref{maxwellwaveeqn}):
\begin{eqnarray}
\left(\frac{\partial^2}{\partial^2x}-\frac{1}{c^2}\frac{\partial^2}{\partial^2t}\right)E_0e^{i(kx-\omega t)}&=&0,\\
\left(-k^2+\frac{\omega^2}{c^2}\right)E_0e^{i(kx-\omega t)}&=&0.
\end{eqnarray}
Solving for the wave vector, we arrive at the dispersion relation
for light in free space:
\begin{equation}
k=\frac{\omega}{c},\label{dispersion1}
\end{equation}
or more familiarly
\begin{equation}
\nu\lambda=c,\label{dispersion2}
\end{equation}
where $c$ is the wave propagation speed, in this case the speed of
light in vacuum. These solutions represent classical electromagnetic
waves, which we know are somehow related to the quantum theory's
photons.

Recall from Einstein and Compton that the energy of a photon is
$\mathcal{E}=h\nu=\hbar\omega$ and the momentum of a photon is
$p=h/\lambda=\hbar k$.  We can rewrite Eq.~(\ref{planewavesoln})
using these relations:
\begin{equation}
E(x,t)=E_0e^{\frac{i}{\hbar}(px-\mathcal{E} t)}.\label{planewavesoln2}
\end{equation}
Substituting this into Eq.~(\ref{maxwellwaveeqn}), we find
\begin{eqnarray}\label{waveequ}
\left(\frac{\partial^2}{\partial^2x}-\frac{1}{c^2}\frac{\partial^2}{\partial^2t}\right)E_0e^{\frac{i}{\hbar}(px-\mathcal{E} t)}&=&0,\label{photon1}\\
-\frac{1}{\hbar^2}\left(p^2-\frac{\mathcal{E}^2}{c^2}\right)E_0e^{\frac{i}{\hbar}(px-\mathcal{E}
t)}&=&0\label{photon2},
\end{eqnarray}
or
\begin{equation}
\mathcal{E}^2=p^2c^2.
\end{equation}
This is just the relativistic total energy,
\begin{equation}
\mathcal{E}^2=p^2c^2+m^2c^4,
\end{equation}
for a particle with zero rest mass, which is reassuring since light
is made of photons, and photons travel at the speed of light in
vacuum, which is only possible for particles of zero rest mass.

We now assume with de Broglie that frequency and energy, and
wavelength and momentum, are related in the same way for classical
particles as for photons, and consider a wave equation for non-zero
rest mass particles. That means, we want to end up with
$\mathcal{E}^2=p^2c^2+m^2c^4$ instead of just
$\mathcal{E}^2=p^2c^2$. Since we do not deal with an electric field
any more, we give the solution to our wave equation a new name, say
$\Psi$, and simply call it the wave function. In doing so, we have exploited that Eq.~(\ref{waveequ}) is homogenous, and hence the units of the function operated upon are arbitrary. Instead of
Eq.~(\ref{photon2}), we would now like
\begin{equation}
-\frac{1}{\hbar^2}\left(p^2-\frac{\mathcal{E}^2}{c^2}+m^2c^2\right)\Psi\,e^{\frac{i}{\hbar}(px-\mathcal{E}
t)}=0,
\end{equation}
which we can get from
\begin{equation}
\left(\frac{\partial^2}{\partial^2x}-\frac{1}{c^2}\frac{\partial^2}{\partial^2t}-\frac{m^2c^2}{\hbar^2}\
\right)\Psi\,e^{\frac{i}{\hbar}(px-\mathcal{E}
t)}=0.\label{kleingordon1D}
\end{equation}
In the discussion of light as a wave or a collection of photons, it
turns out that the square of the electric field is proportional to
the number of photons. By anology, we demand that our wave function,
\begin{equation}
\Psi(x,t)=\Psi_0e^{\frac{i}{\hbar}(px-\mathcal{E} t)},\label{quantsoln}
\end{equation}
be normalizable to unit probability. Then, the probability that the
particle is located somewhere in space,
\begin{equation}
\int_{-\infty}^{\infty}\Psi^*\Psi d\,x=1,
\end{equation}
as it should be.

Removing restriction to one dimension and rearranging, we recognize
this as the Klein-Gordon equation for a free particle,
\begin{equation}
\nabla^2\Psi-\frac{m^2c^2}{\hbar^2}\Psi=\frac{1}{c^2}\frac{\partial^2\Psi}{\partial^2t}.\label{kleingordon}
\end{equation}

The Klein-Gordon equation is a relativistic equation, the
Schr\"{o}dinger equation is not. So to ultimately arrive at the
Schr\"{o}dinger equation, we must make the assumptions necessary to
establish a non-relativistic equation.

The first step in considering the non-relativistic case is to
approximate $\mathcal{E}^2=p^2c^2+m^2c^4$ as follows:
\begin{eqnarray}
\mathcal{E}&=&mc^2\sqrt{1+\frac{p^2}{m^2c^2}}\label{emc},\\
&\approx&mc^2\left(1+\frac{1}{2}\frac{p^2}{m^2c^2}\right),\\
&\approx&mc^2+\frac{p^2}{2m}=mc^2+\mathcal{T}.
\end{eqnarray}
We recognize this last term as the classical kinetic energy,
$\mathcal{T}$. We can then rewrite the wave equation,
Eq.~(\ref{quantsoln}), as
\begin{eqnarray}
\Psi(x,t)&=&\Psi_0e^{\frac{i}{\hbar}(px-mc^2t-\mathcal{T} t)},\\
&=&e^{-\frac{i}{\hbar}mc^2t}\Psi_0e^{\frac{i}{\hbar}(px-\mathcal{T}t)}.\label{QuantSolnNonrel}
\end{eqnarray}

We have assumed that the particle velocity is small such that $mv\ll
mc$, which implies that $p^2\ll m^2c^2$. This means that the leading term in
Eq.~(\ref{QuantSolnNonrel}), $\exp(-imc^2t/\hbar)$, will oscillate
much faster than the last term, $\exp(i\mathcal{T}t/\hbar)$. Taking
advantage of this, we can write
\begin{equation}
\Psi=e^{-\frac{i}{\hbar}mc^2t}\phi,
\end{equation}
where
\begin{equation}
\phi=\Psi_0e^{\frac{i}{\hbar}(px-\mathcal{T}t)}.
\end{equation}
Then
\begin{eqnarray}
\frac{\partial\Psi}{\partial\,t}&=&-\frac{i}{\hbar}mc^2e^{-\frac{i}{\hbar}mc^2t}\phi+e^{\frac{-i}{\hbar}mc^2t}\frac{\partial\phi}{\partial\,t}\\
\frac{\partial^2\Psi}{\partial\,t^2}&=&\Bigl(-\frac{m^2c^4}{\hbar^2}e^{-\frac{i}{\hbar}mc^2t}\phi-\frac{2i}{\hbar}mc^2e^{-\frac{i}{\hbar}mc^2t}\frac{\partial\phi}{\partial\,t}\Bigl)+e^{-\frac{i}{\hbar}mc^2t}\frac{\partial^2\phi}{\partial\,t^2}.
\end{eqnarray}
The first term in brackets is large and the last term is small. We
keep the large terms and discard the small one. Using this
approximation in the Klein-Gordon equation,
Eq.~(\ref{kleingordon1D}), we find
\begin{eqnarray}
e^{-\frac{i}{\hbar}mc^2t}\Bigl[\frac{\partial^2}{\partial\,x^2}+\frac{2im}{\hbar}\frac{\partial}{\partial\,t}\Bigl]\phi&=&0,\\
\frac{\partial^2\phi}{\partial\,x^2}+\frac{2im}{\hbar}\frac{\partial\phi}{\partial\,t}&=&0.
\end{eqnarray}
Again rearranging and generalizing to three spatial dimensions, we
finally arrive at the Schr\"{o}dinger equation for a free particle
(zero potential):
\begin{equation}
-\frac{\hbar^2}{2m}\nabla^2\phi=i\hbar\frac{\partial\phi}{\partial
t},
\end{equation}
where the non-relativistic wave function $\phi$ is also constrained to the condition that it be normalizable to unit probability.

\section{Conclusion}
The simple derivation of the Schr\"{o}dinger equation provided here requires only basic knowledge of the electromagnetic scalar wave equation and the basics of Einstein's special theory of relativity. Both of these topics, and the approximations used in our derivation, are commonly covered prior to a first course in quantum mechanics. Taking the approach that we have outlined exposes students to the reasoning employed by the founders themselves. Though much has been done to refine our understanding of quantum mechanics, taking a step back and thinking through the problem the way they did has merit if our goal as educators is to produce the next generation of Schr\"{o}dingers.

We have glossed over the statistical interpretation of quantum mechanics, which is dealt with superbly in any number of the standard textbooks, and particularly well in Griffiths' text \cite{griffiths}. We have also considered only single particles. Many independent particles are a trivial extension of what we have shown here, but in the presence of interparticle coupling, quantum statistical mechanics is a distraction we can do without, given the narrow objective outlined in the title of this paper. Spin, which is relevant for a fully relativistic treatment of the electron or when more than one particle is considered, has also not been discussed. An obvious next step would be to consider a particle in a potential, but we believe that doing so would result in diminishing returns due to the added complications a potential introduces. What we have shown here is the missing content for the lecture on day one in an introductory quantum mechanics course. Spin, interparticle coupling, and potentials are already adequately covered elsewhere. 

\appendix
\section{The Electromagnetic Wave Equation}

The wave equation governing electromagnetic waves in free space is
derived from Maxwell's equations in free space, which are:
\begin{eqnarray}
\nabla\times\mathbf{E}&=&-\frac{\partial\mathbf{B}}{\partial\,t},\label{Faraday}\\
\nabla\times\mathbf{B}&=&\frac{1}{c^2}\frac{\partial\mathbf{E}}{\partial\,t},\label{Ampere}\\
\nabla\cdot\mathbf{E}&=&0,\label{Gauss}\\
\nabla\cdot\mathbf{B}&=&0,\label{Monopole}
\end{eqnarray}
where $c$ is the speed of light in vacuum, $\mathbf{E}$ is the
electric field, and $\mathbf{B}$ is the magnetic field. The first
equation embodies Faraday's law and illustrates the generation of a
voltage by a changing magnetic field. This equation is the basis of
electric generators, inductors, and transformers. The second
equation embodies Ampere's law and is the magnetic analogy of the
first equation. It explains, for example, why there is a circulating
magnetic field surrounding a wire with electrical current running
through it. It is the basis of electromagnets and the magnetic poles
associated with the rotating ion core in the earth. The last two
equations are embodiments of Gauss' law for electricity and for
magnetism, respectively. In the case of electricity, it is
consistent with Coulomb's law and stipulates that charge is a source
for the electric field. If charges are present, then the right-hand
side of Eq.~(\ref{Gauss}) is non-zero and proportional to the charge
density. The magnetic case is often referred to as the ``no magnetic
monopoles'' law. Since there are no magnetic monopoles (intrinsic
magnetic charge carriers), Eq.~(\ref{Monopole}) always holds.

Applying the curl operator to both sides of Eq.~(\ref{Faraday}) and substituting $\nabla\times\mathbf{B}$ from Eq.~(\ref{Ampere}), we find:
\begin{equation}
\nabla\times(\nabla\times\mathbf{E})=-\frac{1}{c^2}\frac{\partial^2\mathbf{E}}{\partial^2t}\label{waveeqn1}.
\end{equation}
Next, we apply the familiar vector identity\cite{jackson}, $\nabla\times(\nabla\times\Box)=\nabla(\nabla\cdot\Box)-\nabla^2\Box$, where $\Box$ is any vector, to the left hand side of Eq.~(\ref{waveeqn1}):
\begin{equation}
\nabla\times(\nabla\times\mathbf{E})=\nabla(\nabla\cdot\mathbf{E})-\nabla^2\mathbf{E}\label{waveeqn2}.
\end{equation}
From Eq.~(\ref{Gauss}), this reduces to:
\begin{equation}
\nabla^2\mathbf{E}-\frac{1}{c^2}\frac{\partial^2\mathbf{E}}{\partial^2t}=0,\label{waveeqn3}
\end{equation}
which is the electromagnetic wave equation.

\begin{acknowledgements}
The authors would like to thank Jeanette Kurian for being the first undergraduate to test this approach to quantum mechanics on, and Wei Min and Norris Preyer for commenting on the manuscript.
\end{acknowledgements}

\newpage\clearpage

\end{document}